\documentclass[12pt,preprint,epsf]{aastex}

\shorttitle{Sky Brightness at Cerro Tololo}
\shortauthors{Krisciunas et al.} 

\begin{document}
\received{1 September 2009}

\title{Light pollution at high zenith angles, as measured at Cerro 
Tololo Inter-American Observatory\altaffilmark{1}}
\author{
Kevin Krisciunas,\altaffilmark{2}
Hector Bogglio,\altaffilmark{3}
Pedro Sanhueza,\altaffilmark{3}
and Malcolm G. Smith\altaffilmark{4}
}
\altaffiltext{1}{Based in part on observations taken at the Cerro Tololo
Inter-American Observatory, National Optical Astronomy Observatory, 
which is operated by the Association of Universities for Research in 
Astronomy, Inc. (AURA) under cooperative agreement with the National 
Science Foundation.}
\altaffiltext{2}{George P. and Cynthia Woods Mitchell Institute for Fundamental 
Physics \& Astronomy, Texas A. \& M. University, Department of Physics,
  4242 TAMU, College Station, TX 77843; {krisciunas@physics.tamu.edu} }
\altaffiltext{3}{Oficina de Protecci\'{o}n de la Calidad del Cielo del Norte de Chile (OPCC),
1606 Cisternas, La Serena, Chile; {hbogglio@opcc.cl}, {psanhueza@opcc.cl} }
\altaffiltext{4}{Cerro Tololo Inter-American Observatory, Casilla 603,
  La Serena, Chile; {msmith@ctio.noao.edu} }

\begin{abstract} 

On the basis of measurements of the $V$-band sky brightness obtained at Cerro
Tololo Inter-American Observatory in December 2006 and December 2008 we confirm the
functional form of the basic model of \citet{Gar89,Gar91}.  At high zenith angles
we measure an enhancement of a factor of two over Garstang's later model when there
is no marine cloud layer over La Serena/Coquimbo.  No corresponding enhancement
is found in the $B$-band.

\end{abstract}


\section{Introduction}

In a previous paper \citep[][hereafter K07]{Kri_etal07} we presented measurements 
of the $BVRI$ sky brightness at Cerro Tololo Inter-American Observatory, obtained 
with the CTIO 0.9-m, 1.3-m, and 1.5-m telescopes and CCD detectors.  As a sanity 
check we also obtained $B$- and $V$-band data with a 15-cm reflector and single 
channel photometer previously used by \citet{Kri97} for an 11-year series of sky 
brightness measurements obtained at the 2800-m level of Mauna Kea.  We found that 
at a zenith angle of $\sim$80 deg we could measure the effect of artificial 
lights in La Serena some 55 km away.  The excess $V$-band light amounted to an 
enhancement of 44 to 72 percent above the model of \citet{Gar89}.  The December 
2006 single channel $B$- and $V$-band measuremens are included in the analysis of 
the present paper.

In December of 2008 we obtained more data with the single channel system of 
\citet{Kri96}. Our purpose was twofold: 1) to confirm the functional form of the 
basic model of Garstang by taking data at the full range of zenith angles; and 2) 
to determine a baseline level of light pollution measurable from Cerro Tololo.  
As La Serena and Coquimbo's population continues to grow, we can use these 
measurements to determine the effectiveness of the lighting control measures in 
place.

\section{The Data}

Our more and more antiquated, but still functional, photometric system consists of a
15-cm f/5.9 Newtonian reflector and a very light photometer whose light sensitive
element is an uncooled RCA 931A photomultiplier (pm) tube $-$ effectively the
same as a 1P21 tube.  The $V$-band filter is a 2 mm thick piece of Schott GG495
glass, which has a short wavelength cut on at about 4800 \AA.  At 5500 \AA\ it
reaches a maximum transmission of 91 percent, which it retains to red end of the
visible portion of the electromagnetic spectrum. The {\em effective} $V$-band filter 
transmission results from the 
combination of this transmission curve and the quantum efficiency (QE) of the pm tube,
which falls to near zero at 7000 \AA \citep[][Fig. 1]{lal62}.
The $B$-band filter is composed of a 2 mm thick piece of Schott GG385 glass and
a 1 mm thick piece of Schott BG-12 glass.  Together they give a filter transmission
which cuts on at 3600 \AA, has peak transmission at 4000 \AA, diminishes to
zero shortly after 5000 \AA, but then has a $\sim$ 15 percent red
leak starting at 7000 \AA.  Given the near zero QE of the tube at the red end,
the red leak would have no significant effect on the $B$-band measurements.

From 28 to 30 December 2008 UT our typical observing strategy was as follows: 1) 
start and end the night with measurements of standard stars chosen from the {\em 
Bright Star Catalogue} \citep{Hof_Jas82} so that the photometric zeropoints are 
interpolated and never extrapolated; 2) measure the background sky near the 
zenith, then very low in the sky, then at an intermediate zenith angle; 3) 
measure a standard star every 40 minutes or so.  Our principal standard stars 
were BS 1179 and $\zeta$ Cae, with $\rho$ For observed less often.  The first two 
stars could be observed on the same gain setting used for the sky brightness 
measures.  This eliminates one source of potential systematic error. The output 
of the pm tube is amplified by a factor up to 10$^6$ with an analog amplifier. 
The output is displayed real-time on a milliammeter and recorded with a Hewlett 
Packard strip chart recorder.

To determine the photometric zeropoints each night we assumed mean extinction 
coefficients of $k_v = 0.154$ mag/airmass and $k_b$ = 0.274 mag/airmass, based on 
several years of CCD photometry at CTIO.  Since our standards were observed 
between 1.0 and 1.1 airmasses, the adopted extinction values hardly mattered, but 
on any given night the derived sky brightness values are systematically in error 
by the difference of the true extinctions and the assumed values. Of greater 
consequence is the drift of the zeropoints with time.  We found that the drift 
was linear and amounted to a maximum of 0.03 mag/hour.  The sky brightness 
measures presented here take into account these linear drifts.  It should be 
noted that extinction corrections are not made to sky brightness measures, only 
to measures of standard stars.

Once we determine the equivalent $B$- and $V$-band magnitudes of the sky in our 
6.522 square arc minute beam, we obtain the number of magnitudes per square arc 
second by adding 10.927 to the single-beam magnitudes.  This additive factor is 
just 2.5 log$_{10} A$, where $A$ is the beam size in square arc seconds.

Almost all of our high zenith angle measurements made in December 2006 and 
December 2008 were taken over La Serena.  On nights when there is no marine cloud 
layer, the city is clearly visible down below.  In Fig. \ref{la_serena} we show a 
photo of this situation, obtained by one of us (MGS).

The sky-brightness data are found in Table \ref{data}.  For the $V$-band we also give
the flux in nanoLamberts, using Eq. 3 of K07 or its equivalent \citep[][Eq. 27]{Gar89}.

Fluxes in nanoLamberts are technically only relevant for the $V$-band, as it 
approximates the response of the human eye.  We can convert the $B$-band sky 
brightness in mag/sec$^2$ to the flux $b_B$ in photons/cm$^2$/sec/steradian by 
using the inverse of Eq. 39 of \citet{Gar89}, namely:

\begin{displaymath}
b_B \; = \; {\rm exp} \left[\frac{41.965 - B}{1.0857}\right] \; .
\end{displaymath}

The observed flux of the sky near the zenith can effectively be corrected to the 
zenith by dividing it by the secant of the zenith angle.  This is basically the 
same as Eq. 19 of \citet{Sch90}:

\begin{displaymath}
B_{zen} \; = \; B_{obs} / (1 \; + \; Z^2_{rad}/2) \; \; ,
\end{displaymath}

\parindent = 0 mm

where $Z_{rad}$ is the zenith angle in radians.

\parindent = 9 mm

\section{Discussion}

The mean sky brightness at the zenith in December 2008 was $V$ = 21.958 
mag/sec$^2$.  An individual $V$-band measurement at that flux level was accurate 
to $\pm$ 0.060 mag/sec$^2$. This is based on 17 observations within 27 degrees of 
the zenith and corrected to the zenith using the formula of \citet{Sch90} given 
above.  The mean zenith $V$-band flux was 56.24 nL, with a 1-$\sigma$ scatter of 
$\pm$ 3.11 nL (5.5 percent). The mean $B$-band zenith sky brightness was 22.826 
mag/sec$^2$.  An individual $B$-band measurement at that flux level was accurate 
to $\pm$ 0.10 mag/sec$^2$. These mean values are amongst the faintest levels we 
have ever measured, indicating that the solar minimum was still in effect in 
December 2008.
 
The individual $V$-band data points from December 2006 and December 2008 are 
shown in Fig.  \ref{vband}. We have included the locus of \citet{Gar89} derived 
for Mt. Graham, offset by +3.0 nL, and the locus of \citet{Gar91} for Junipero 
Serra (California), similarly offset.  Since we did not correct our measurements 
for the presence of faint stars in the beam, some offset is understandable.
Garstang's improved 1991 model includes the addition of an ozone layer, a more
accurate representation of the atmospheric molecular density variation as a 
function of height, a better mathematical representation of the scattering 
angular function of aerosols, and a thin layer of dust of arbitrary optical 
thickness and height above sea level.  These modifications lead to a reduction
of the predicted brightness of the night sky.

The {\em ratio} of our $V$-band data to the slightly offset locus of 
\citet{Gar91} is shown in Fig. \ref{v_ratio}.  This shows that {\em except} at 
high zenith angle we do not measure any significant enhancement of the $V$-band 
sky brightness at CTIO, even when we can clearly see La Serena down below.  For 
comparison the reader is directed to the interesting article by 
\citet{Lug_etal09}, which includes pictorial and graphical data along with 
modeling based on the \citet{Gar91} model and an approximate correction for 
double scattering.

Our $B$-band data, converted to photons/cm$^2$/sec/steradian, are shown in Fig. 
\ref{bband}. Unlike our $V$-band data, there is no strong enhancement of the 
$B$-band light at high zenith angle when the lights of La Serena/Coquimbo are 
visible to the naked eye.  This may be because the artificial lights presently in 
operation do not put out much light in the $B$-band.  Spectra of the scattered 
light should be obtained to check this.

There is one curious ``anomaly'' in Fig. \ref{vband}.  Our final high zenith 
angle measurements on 30 December 2008 were obtained over the town of Andacollo 
$-$ a much more southerly azimuth than La Serena.  Though Andacollo was clearly 
visible (i.e. not covered by cloud), we measured no enhancement of the $V$-band 
flux in this direction compared to the model of \citet{Gar91}.  We surmise that 
this is due to the smaller population in Andacollo and a smaller amount of 
airborne dust there.  A combination of dust and light emitted above the 
horizontal gives us enhanced light levels over La Serena and Coquimbo when they 
are visible at Cerro Tololo.

As the population of La Serena and Coquimbo continues to grow, the observations 
presented here may serve as a reference.  If the lighting ordinances in place are 
effective, we may see only minimally increased light pollution at CTIO.

\vspace {1 cm}

\acknowledgments

We thank Oscar Saa for technical support at the mountain.

\newpage

\begin{deluxetable}{ccccccr}
\tablewidth{0pc}
\tablecaption{CTIO Sky Brightness Values\tablenotemark{a}\label{data}}
\tablehead{   \colhead{UT Date} & \colhead{$\langle$UT$\rangle$} &
\colhead{Azimuth} & \colhead{Zenith Angle} & 
\colhead{Filter} & \colhead{Sky Brightness} & Flux (nL)  } 
\startdata

Dec 28 &  1:41 &  94.75 &  17.81 & V & 22.014 &   53.33  \\
Dec 28 &  2:22 & 268.80 &   4.01 & V & 21.949 &   56.62  \\
Dec 28 &  2:30 & 309.34 &  83.35 & V & 20.626 &  191.50  \\
Dec 28 &  3:11 & 265.72 &  14.61 & V & 21.922 &   58.04  \\
Dec 28 &  3:21 & 271.50 &  69.67 & V & 21.331 &  100.04  \\
Dec 28 &  3:42 & 263.72 &  21.30 & V & 21.926 &   57.83  \\
Dec 28 &  4:03 & 266.87 &  12.95 & V & 21.929 &   57.67  \\
Dec 28 &  4:11 & 309.20 &  83.09 & V & 20.665 &  184.74  \\
Dec 28 &  4:34 & 264.66 &  19.65 & V & 21.917 &   58.31  \\
Dec 28 &  4:41 & 304.36 &  82.36 & V & 20.631 &  190.62  \\
Dec 28 &  4:57 & 263.07 &  24.61 & V & 21.912 &   58.58  \\
\\
Dec 28 &  1:44 &  94.54 &  17.17 & B & 22.781 &   \ldots \\
Dec 28 &  2:18 & 269.04 &   3.14 & B & 22.878 &   \ldots \\
Dec 28 &  2:33 & 308.92 &  83.85 & B & 22.176 &   \ldots \\
Dec 28 &  3:10 & 265.97 &  13.75 & B & 22.821 &   \ldots \\
Dec 28 &  3:25 & 270.99 &  70.54 & B & 22.396 &   \ldots \\
Dec 28 &  3:40 & 263.85 &  20.87 & B & 22.919 &   \ldots \\
Dec 28 &  4:01 & 267.02 &  12.52 & B & 22.829 &   \ldots \\
Dec 28 &  4:09 & 309.49 &  82.76 & B & 22.258 &   \ldots \\
Dec 28 &  4:33 & 264.73 &  19.44 & B & 22.713 &   \ldots \\
Dec 28 &  4:43 & 304.36 &  82.36 & B & 22.205 &   \ldots \\
Dec 28 &  4:55 & 263.21 &  24.18 & B & 22.788 &   \ldots \\
\\
Dec 29 &  3:05 & 265.85 &  14.17 & V & 21.828 &   63.29  \\
Dec 29 &  3:13 & 263.92 &  59.38 & V & 21.616 &   76.94  \\
Dec 29 &  3:24 & 306.93 &  79.19 & V & 21.083 &  125.71  \\
Dec 29 &  3:46 & 263.13 &  23.23 & V & 21.962 &   55.95  \\
Dec 29 &  3:52 & 270.59 &  55.86 & V & 21.712 &   70.43  \\
Dec 29 &  4:01 & 310.08 &  82.10 & V & 21.140 &  119.28  \\
Dec 29 &  4:25 & 265.00 &  18.56 & V & 21.877 &   60.50  \\
Dec 29 &  4:35 & 265.47 &  64.96 & V & 21.667 &   73.41  \\
Dec 29 &  4:40 & 302.73 &  82.53 & V & 21.124 &  121.05  \\
Dec 29 &  4:54 & 263.00 &  24.81 & V & 21.829 &   63.24  \\
\\
Dec 29 &  3:06 & 265.62 &  14.38 & B & 22.740 &  \ldots  \\ 
Dec 29 &  3:15 & 263.69 &  59.82 & B & 22.626 &  \ldots  \\
Dec 29 &  3:26 & 306.63 &  79.54 & B & 22.429 &  \ldots  \\
Dec 29 &  3:45 & 263.26 &  22.80 & B & 22.849 &  \ldots  \\
Dec 29 &  3:55 & 270.34 &  56.30 & B & 22.688 &  \ldots  \\
Dec 29 &  4:03 & 309.79 &  82.42 & B & 22.641 &  \ldots  \\
Dec 29 &  4:27 & 264.87 &  18.99 & B & 22.809 &  \ldots  \\
Dec 29 &  4:33 & 265.70 &  64.53 & B & 22.697 &  \ldots  \\
Dec 29 &  4:43 & 302.31 &  83.08 & B & 22.523 &  \ldots  \\
Dec 29 &  4:52 & 263.14 &  24.38 & B & 22.826 &  \ldots  \\
\\
Dec 30 &  2:04 &  92.50 &  11.14 & V & 21.825 &   63.47  \\
Dec 30 &  2:12 & 309.81 &  81.66 & V & 20.920 &  146.07  \\
Dec 30 &  2:25 & 271.33 &  49.84 & V & 21.670 &   73.21  \\
Dec 30 &  2:51 & 266.49 &  11.99 & V & 21.875 &   60.61  \\
Dec 30 &  2:57 & 267.35 &  56.77 & V & 21.552 &   81.61  \\
Dec 30 &  3:06 & 309.10 &  76.79 & V & 21.203 &  112.55  \\
Dec 30 &  3:34 & 263.73 &  21.27 & V & 21.942 &   56.99  \\
Dec 30 &  3:40 & 271.76 &  53.90 & V & 21.653 &   74.36  \\
Dec 30 &  3:47 & 312.57 &  80.47 & V & 21.289 &  103.98  \\
Dec 30 &  4:02 & 266.43 &  14.22 & V & 21.952 &   56.46  \\
Dec 30 &  4:07 & 268.36 &  59.75 & V & 21.608 &   77.51  \\
Dec 30 &  4:16 & 305.66 &  78.93 & V & 21.160 &  117.10  \\
Dec 30 &  4:34 & 264.11 &  21.35 & V & 21.843 &   62.43  \\
Dec 30 &  4:41 & 262.08 &  82.46 & V & 21.283 &  104.56   \\
Dec 30 &  4:48 & 272.95 &  56.59 & V & 21.711 &   70.50  \\
Dec 30 &  5:00 & 262.32 &  26.94 & V & 21.828 &   63.29  \\
\\
Dec 30 &  2:02 &  92.65 &  11.57 & B & 22.557 & \ldots \\
Dec 30 &  2:16 & 309.22 &  82.32 & B & 22.167 & \ldots \\
Dec 30 &  2:22 & 271.72 &  49.19 & B & 22.511 & \ldots \\
Dec 30 &  2:52 & 266.43 &  12.21 & B & 22.677 & \ldots \\
Dec 30 &  3:00 & 266.99 &  57.42 & B & 22.477 & \ldots \\
Dec 30 &  3:10 & 308.47 &  77.47 & B & 22.373 & \ldots \\
Dec 30 &  3:29 & 264.05 &  20.20 & B & 22.773 & \ldots \\
Dec 30 &  3:41 & 271.63 &  54.11 & B & 22.605 & \ldots \\
Dec 30 &  3:45 & 312.87 &  80.15 & B & 22.327 & \ldots \\
Dec 30 &  3:59 & 266.58 &  13.79 & B & 22.842 & \ldots \\
Dec 30 &  4:09 & 268.11 &  60.18 & B & 22.619 & \ldots \\
Dec 30 &  4:14 & 305.96 &  78.58 & B & 22.289 & \ldots \\
Dec 30 &  4:32 & 264.25 &  20.92 & B & 22.713 & \ldots \\
Dec 30 &  4:43 & 261.83 &  82.89 & B & 22.457 & \ldots \\
Dec 30 &  4:51 & 272.56 &  57.24 & B & 22.500 & \ldots \\
Dec 30 &  4:59 & 262.46 &  26.52 & B & 22.609 & \ldots \\
\enddata
\tablenotetext{a}{Year is 2008.  UT is in hours and minutes.  Azimuth and zenith angle are
in degrees, and are probably accurate to $\pm$ 0.2 deg.  Sky brightness is measured in 
magnitudes per square arc second.  For $V$-band
values the sky brightness is also converted to flux in nanoLamberts.}
\end{deluxetable}

\clearpage

\figcaption[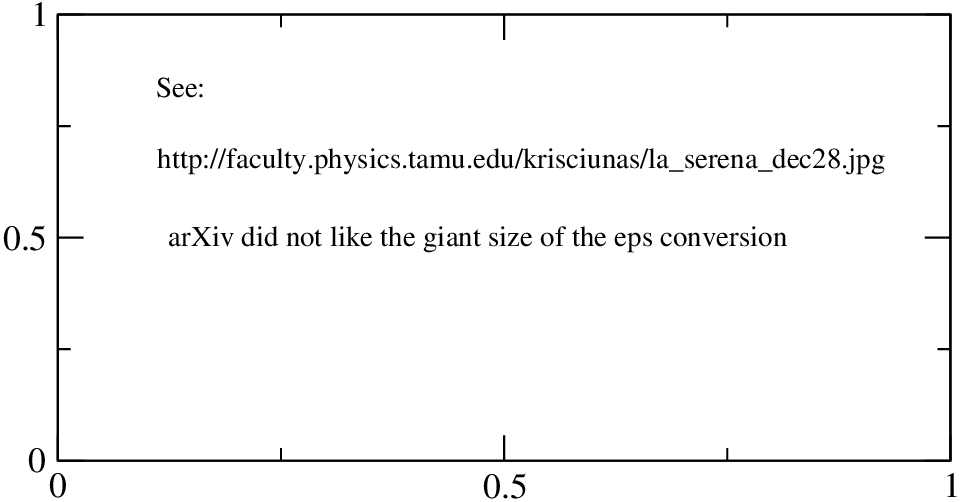]
{15 minute exposure over La Serena and Coquimbo, 28 December 2008 UT, taken
just south of the \#1 40-cm dome at CTIO.
\label{la_serena}
}

\figcaption[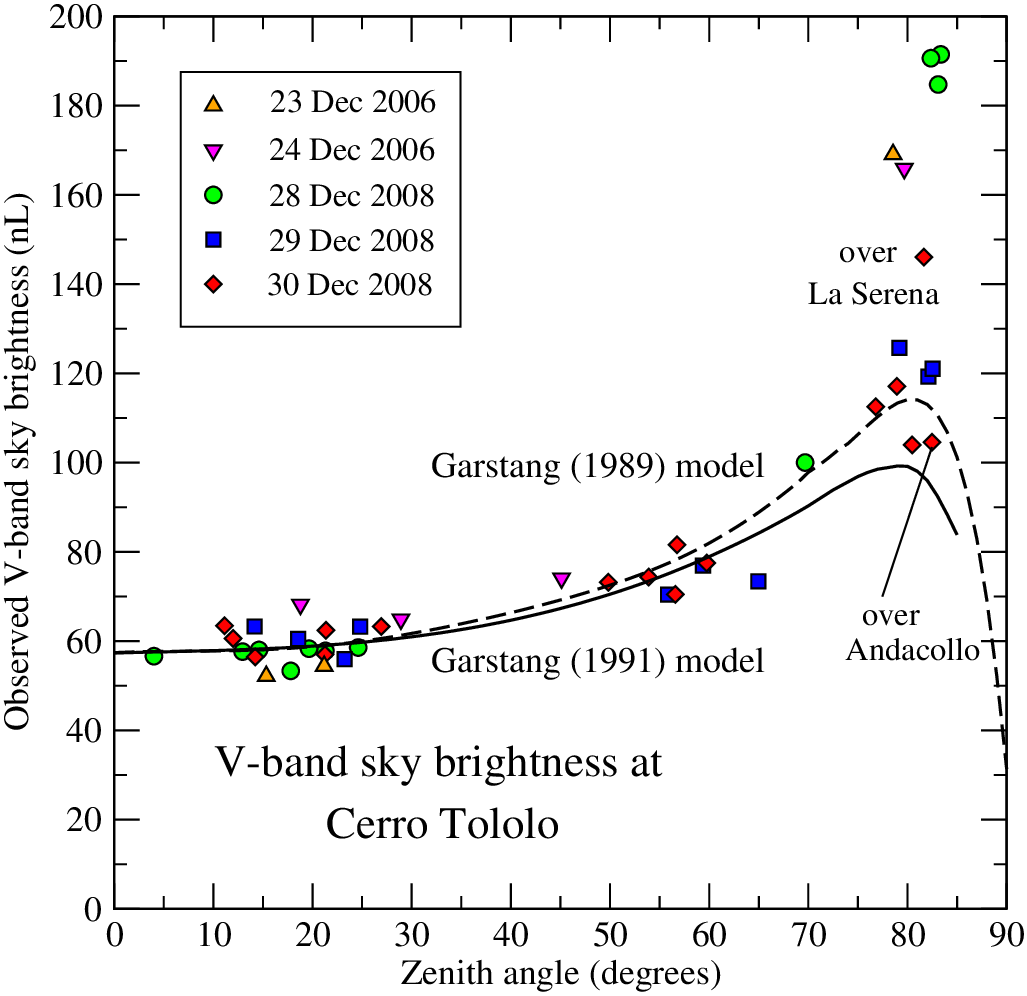]
{Observed V-band sky brightness at Cerro Tololo during solar minimum and at a variety
of zenith angles. The upper (dashed) locus is based on the model of \citet{Gar89} for Mt. Graham.
The lower (solid) locus is based on the model of \citet{Gar91} for Junipero Serra, California.
\label{vband}
}

\figcaption[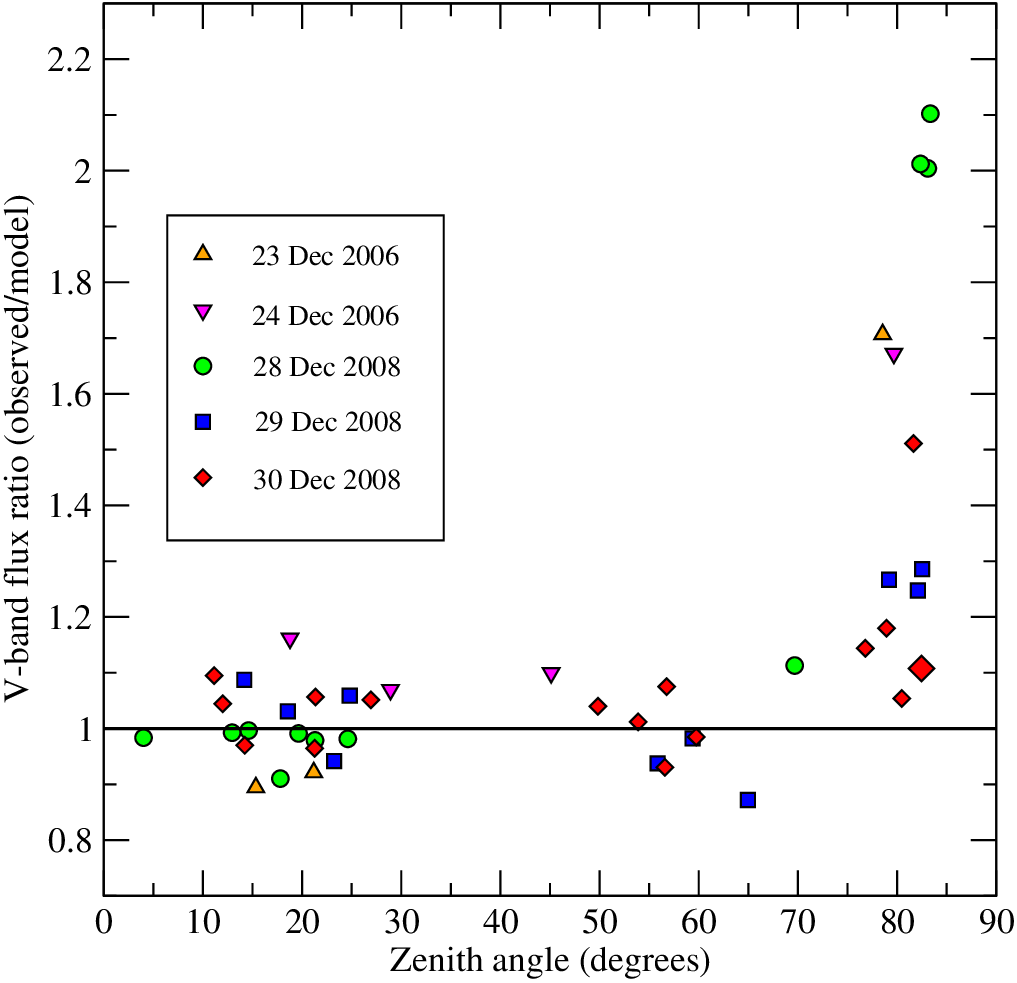]
{Ratio of observed V-band sky brightness and \citet{Gar91}
model shown in Fig. \ref{vband}.  Essentially, only at high zenith angles do we find 
observed sky brightness in excess of Garstang's model.
\label{v_ratio}
}

\figcaption[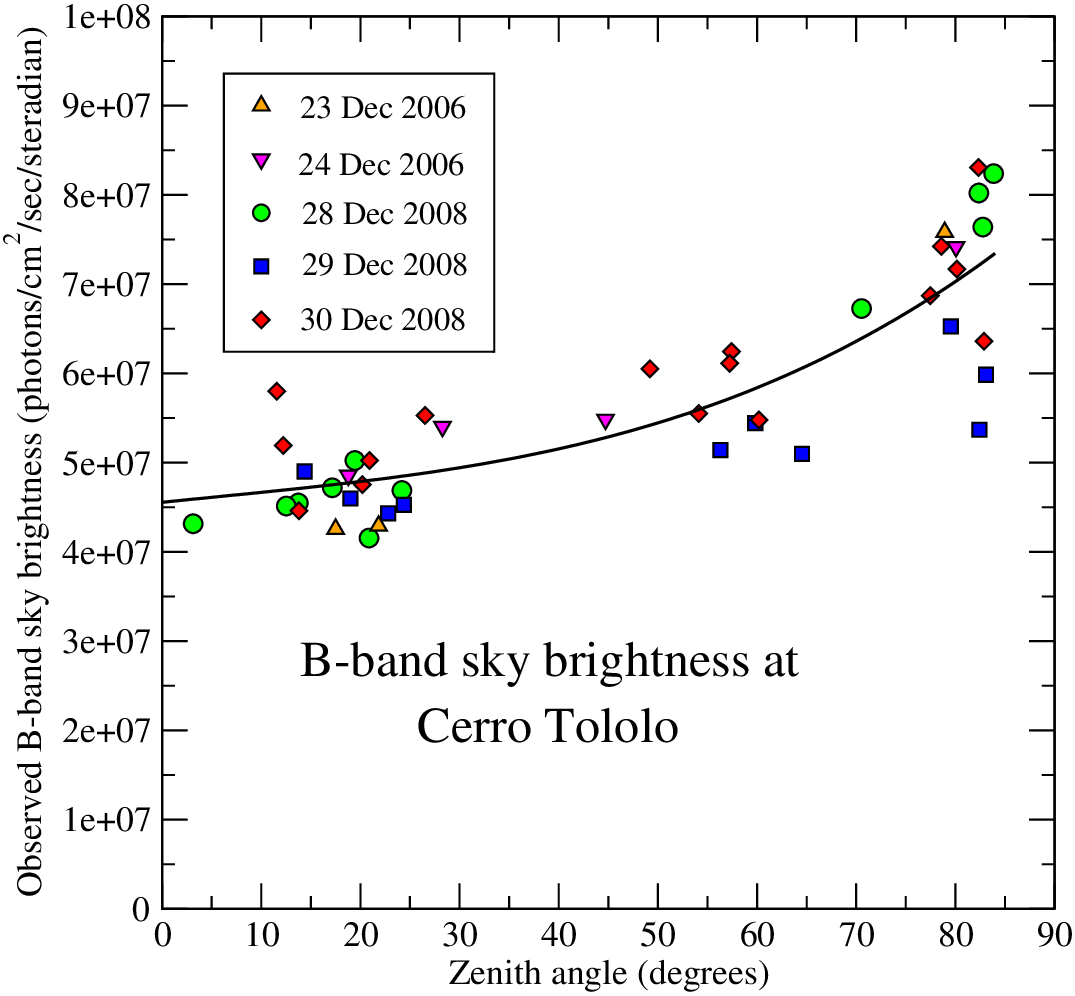]
{Observed B-band sky brightness at Cerro Tololo, converted to flux in cgs
units. A third order polynomial fit is also shown.  \label{bband}
}

\clearpage

\begin{figure}
\plotone{la_serena_dec28.eps}
{\center Krisciunas {\it et al.} Fig. \ref{la_serena}}
\end{figure}

\begin{figure}
\plotone{bobs_v.eps}
{\center Krisciunas {\it et al.} Fig. \ref{vband}}
\end{figure}

\begin{figure}
\plotone{ratio.eps}
{\center Krisciunas {\it et al.} Fig. \ref{v_ratio}}
\end{figure}

\begin{figure}
\plotone{bobs_b.eps}
{\center Krisciunas {\it et al.} Fig. \ref{bband}}
\end{figure}

\end{document}